\documentclass[a4paper,fleqn]{cas-dc}

\usepackage[sort&compress,numbers]{natbib}

\usepackage{color,soul} 

\usepackage{amsmath}
\usepackage{nccmath}

\def\tsc#1{\csdef{#1}{\textsc{\lowercase{#1}}\xspace}}
\tsc{WGM}
\tsc{QE}
\tsc{EP}
\tsc{PMS}
\tsc{BEC}
\tsc{DE}

\begin{document}
\let\WriteBookmarks\relax
\def\floatpagepagefraction{1}
\def\textpagefraction{.001}
\shorttitle{Real-time in water tritium monitor}
\shortauthors{C.D.R. Azevedo et~al.}

\title [mode = title]{Simulation results of a real-time in water tritium monitor }                      
\tnotemark[1]

\tnotetext[1]{This document is the results of the research
   project funded by INTERREG-SUDOE program through the project TRITIUM - SOE1/P4/E0214.}


\author[1]{C.D.R.~Azevedo}[orcid=0000-0002-0012-9918]
\cormark[1]
\ead{cdazevedo@ua.pt}

\author[2]{A.~Baeza}[]
\author[3]{E.~Chauveau}[]
\author[2]{J.A.~Corbacho}[]
\author[4]{J.~D\'iaz}[]
\author[3]{J.~Domange}[]
\author[3]{C.~Marquet}[]
\author[4]{M.~Mart\'inez-Roig}[]
\author[3]{F.~Piquemal}[]
\author[1]{J.F.C.A.~Veloso}[]
\author[4]{N.~Yahlali}[] 

\address[1]{I3N, Physics Department, University of Aveiro, Campus Universit\'ario de Santiago, 3810-193 Aveiro, Portugal}
\address[2]{Universidad de Extremadura, Laboratorio de Radioactividad Ambiental. Servicio de apoyo a la investigaci\'on, C\'aceres, Spain}
\address[3]{Universit\'e de Bordeaux and CNRS, CENBG, Gradignan Cedex, France}
\address[4]{Instituto de F\'isica Corpuscular, Centro mixto CSIC-Universidad de Valencia, Paterna, Spain}

\cortext[cor1]{Corresponding author}

\begin{abstract}
In this work we present simulation results for a modular tritium in-water real-time monitor. The system allows for scalability in order to achieve the required sensitivity. The modules are composed by 340 uncladed scintillating fibers immersed in water and 2 photosensors for light readout. Light yield and Birks' coefficient uncertainties for low energy beta particles is discussed. A study of the detection efficiency according to the fiber length is presented. Discussion on the system requirements and background mitigation for a device with sensitivity of 100\,Bq/L, required to comply with the European directive 2013/51/Euratom, is presented. Due to the low energetic beta emission from tritium a detection efficiency close to 3.3\% was calculated for a single 2\,mm round fiber.

\end{abstract}
%
%

\maketitle

\section{Introduction}
Tritium, a hydrogen isotope, is produced by neutron capture in nuclear power plants mainly in the water coolant and moderator, in quantities depending on the reactor type, i.e., light-water reactors (LWR), pressurized water reactors (PWR), pressurized heavy water reactor (PHWR), etc. The tritium produced in the coolant is partially or totally released in the air or in the water and in the latter case the dominant form is through tritiated water (HTO or T\textsubscript{2}O) \cite{Hou2018}.
Tritium has a half-life of 12.3 years decaying by $\beta$ emission with average energy of 5.7\,keV and maximum of 18\,keV. Despite the low energy $\beta$ emission which cannot penetrate the skin, several state regulations have been released in order to regulate the maximum quantity of tritium in drinking water. For example, the U.S. Environmental Protection Agency (EPA) that sets a maximum of 740\,Bq/L \cite{USNRC} or the E.U. Council Directive 2013/51/Euratom which establishes the limit to 100\,Bq/L \cite{directive2013}.
For such low required limits the mostly used and suitable technique is the Liquid Scintillation Counting (LSC) which besides producing toxic waste also requires sample collection, preparation and analysis which takes at least 2 days. Real-time monitors based on solid scintillators have been implemented and reported in several published works \cite{Hofstetter1992,Hofstetter1995,Rathnakaran2000} to have detection limits of the order of dozens of kBq/L. This is 2 orders of magnitude higher than the limit required by the E.U. directive. In 1999 J.W. Berthold and L.A. Jeffers \cite{Berthold1999} proposed the use of scintillating fibers with 2\% of fluor doped fiber clad. Unfortunately the authors concluded that their system was unable to comply with the U.S. EPA required sensitivity, finding in addition several issues related to the stability of fluor-doped clad immersed in water.\\
In this work the simulation studies of a real-time tritium monitor based on uncladed scintillation fibers is presented. This monitor was developed by the INTERREG SUDOE TRITIUM Project \cite{tritiumsudoe, Azevedo2018} aiming to achieve the 100\,Bq/L sensitivity required by the E.U. directive.

\section{The setup}
\begin{figure}
	\centering
		\includegraphics[width=\columnwidth]{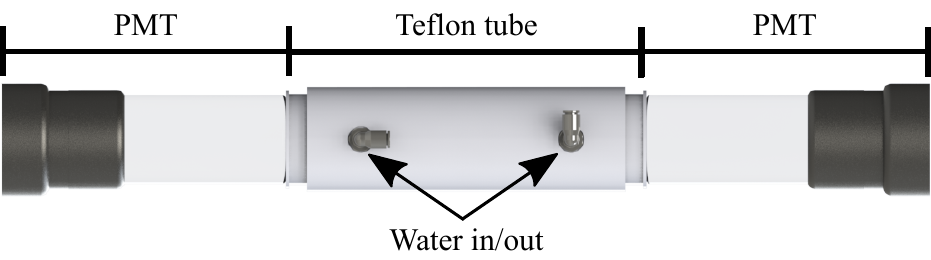}
	\caption{The TRITIUM-1 prototype studied and developed in this work}
	\label{FIG:1}
\end{figure}

In order to measure the tritium activity in water, a detector based on a modular and scalable design was conceived. Each module, represented in Figure \ref{FIG:1}, is composed by a teflon tube (PTFE),  from now on denominated as "cell", which contains scintillating fibers immersed in water. The teflon material was chosen in order to maximize the optical photon reflection on the inner surface of the tube. Also, its mechanical properties allows to seal it against acrylic (PMMA) disks by just applying pressure radially on the external walls of the tube using clamps. The PMMA disks (1\,cm thick) are used as optical windows and to water-tight the cell. The cell also contains water inlet and outlet in order to keep a constant water flux.
Inside the cell, 2\,mm diameter round fibers (Saint-Gobain Crystals BCF-10) \cite{SaintGobain}, specially ordered uncladed, are placed. The clad would prevent the use of regular scintillating fibers for tritium detection, as its $\beta$ emission is not energetic enough to reach the fiber core. In the eventual cases that it occurs, the energy of the particle would be strongly decreased reducing the light produced by the fiber. 
The 2\,mm diameter fibers were chosen to maximize the detection area in contact with water. Lower fiber diameters could also be used allowing a higher fill density of the detector vessel. At the moment we cannot guaranty an uniform or efficient water flux between the fibers in case of high packaging densities.
To maximize the light collection large-area photodetectors were chosen in order to collect not only the light guided by the fibers but also the photons traveling in the water: 2" Hamamatsu R2154-02 PMTs selected for their QE and availability.  In order to discard false-events due to the PMTs photocathode self-emission, each module uses 2 PMTs placed at each end of the cell in coincidence mode. Finally, the PMTs are optically coupled to the PMMA windows using optical grease (Saint-Gobain BC630). At present an analogous system using SiPMs is being studied. PMTs offer advantage over SiPMs in price (for the same sensitive area), lower dark-counts and lower number of electronic channels which simplifies the electronics requirements and correspondent price. On the other side, SiPMs present higher quantum efficiency for the fibers scintillation wavelength, operation at low voltage and they are compact which allows for a higher number of modules in the same space.

\section{Simulation details}
\subsection{Physics list and materials properties}
\label{section:physics}

\begin{sloppypar}
For the calculations presented here, the GEANT4 \mbox{toolkit \cite{GEANT4}} was used. In order to comply with the low energy calculations requirement the Livermore physics list (\emph{G4EmLivermorePhysics}) was used, which includes electromagnetic processes as Bremsstrahlung, Coulomb scattering, atomic de-excitation (fluorescence) and other related effects. The fiber material properties (polystyrene) was taken from \emph{GEANT4 NIST} database (\emph{G4\_POLYSTYRENE}). The fiber emission spectrum and the light attenuation were taken from the manufacturer data \cite{SaintGobain}. The water optical properties were taken from Ref.\cite{Buiteveld1994} while the PMMA (optical windows), teflon and borosilicate glass (PMT windows) optical data were taken from Ref.\cite{Argyriades2011}. A refraction index of 1.46 was considered for the silicon optical grease. The typical quantum efficiency of the PMTs was taken from the manufacturer's datasheet \cite{R2154}.
\end{sloppypar}

\subsection{Source simulation}
\label{source} 
   \begin{figure}
	\centering
		\includegraphics[width=\columnwidth]{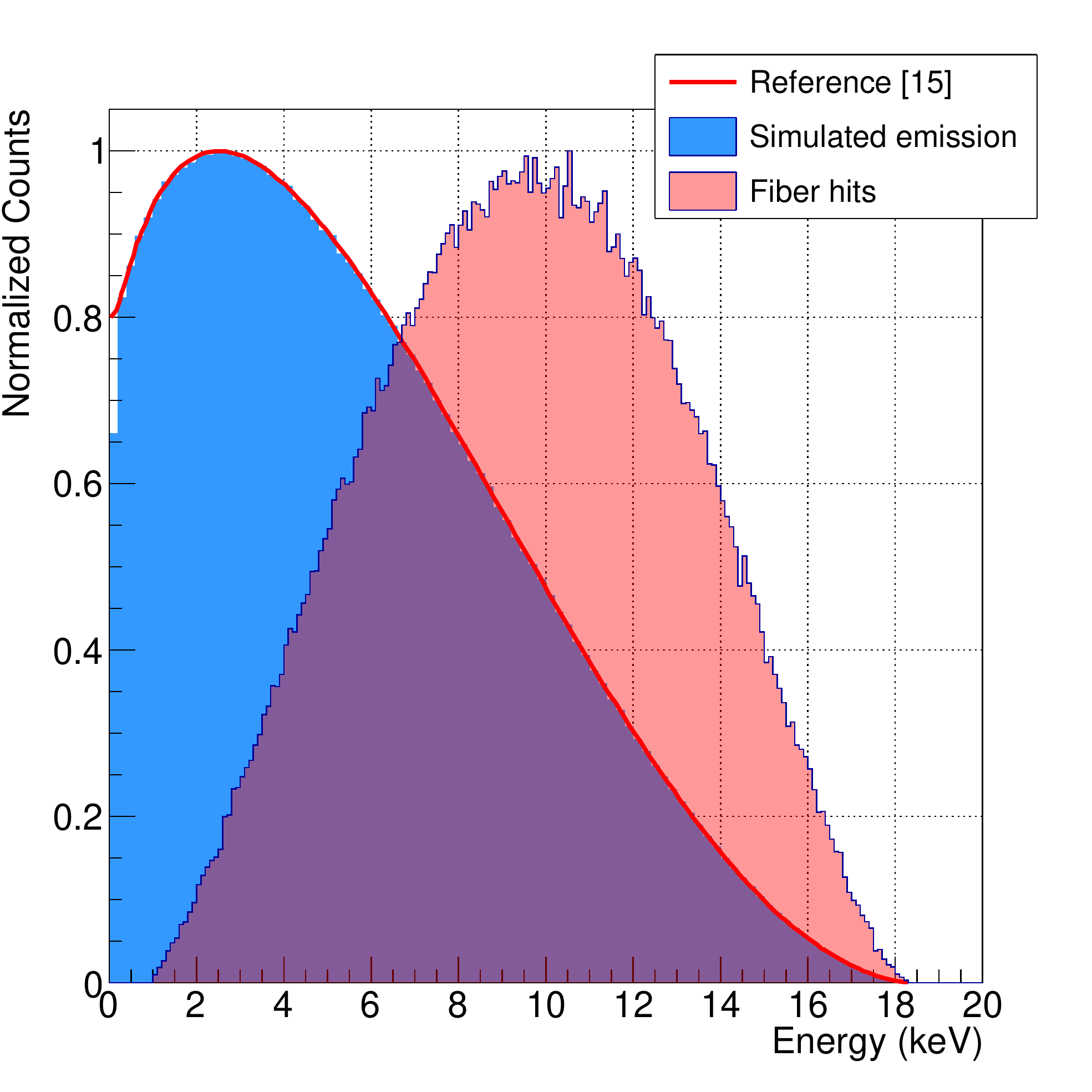}
	\caption{Tritium beta energy distribution (blue) generated with GEANT4 by sampling the distribution obtained from \cite{Mertens2015}. The red distribution  corresponds to the initial energy of the electrons that deposit energy in the fiber. Each distribution was normalized to its maximum for visualization proposes.}
	\label{FIG:2}
\end{figure}

\begin{figure}
	\centering
		\includegraphics[width=\columnwidth]{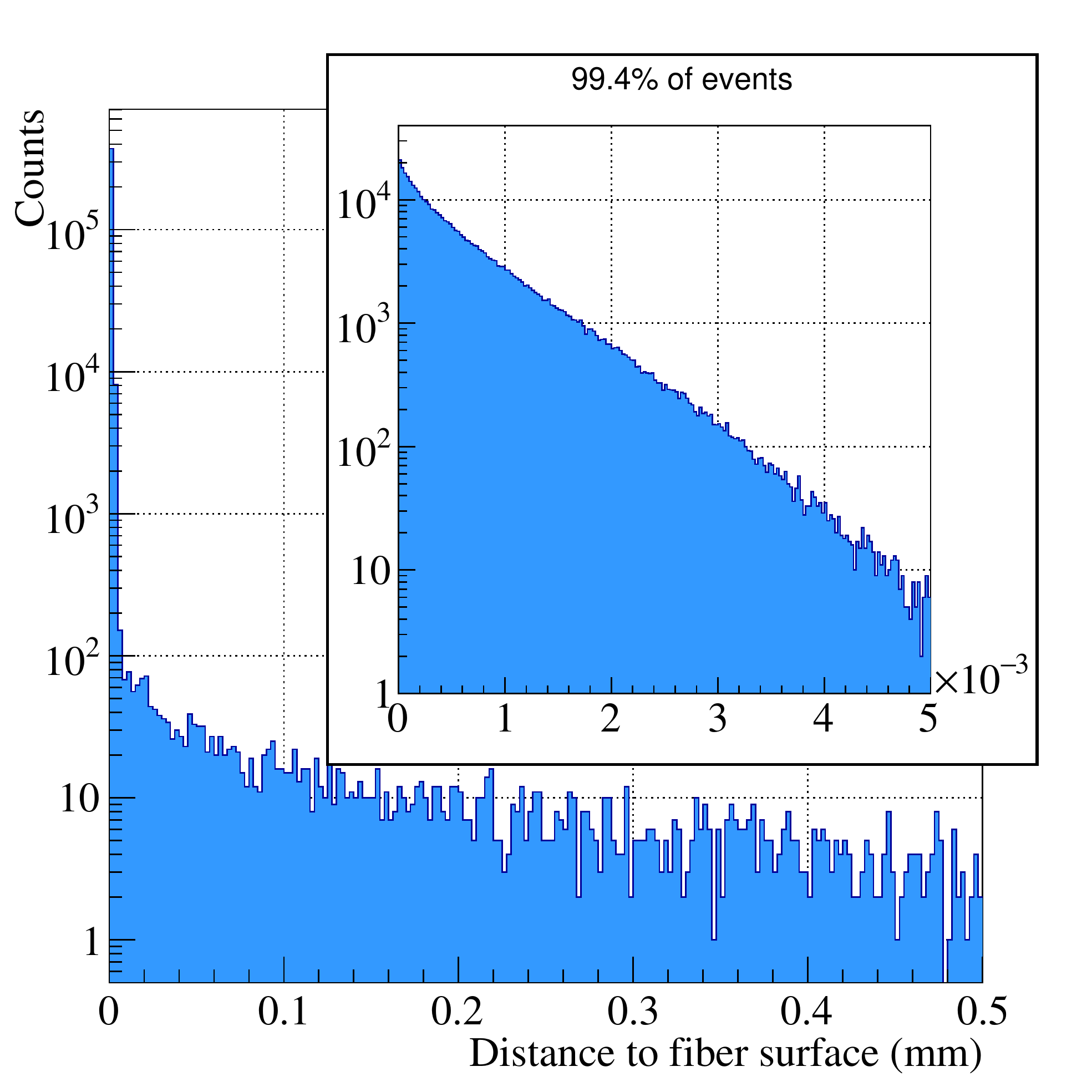}
	\caption{Histogram of the distances between the $\beta$ emission position and fiber surface for detected electrons. The box is a detail containing the re-binned data below 5\,$\mu$m}
	\label{FIG:3b}
\end{figure}

 In this section an \emph{ad-hoc} study on the tritium decay and the interaction with a single fiber is carried out. The $\beta$ source  for the tritiated water was considered as a water tube with internal diameter equal to the fiber diameter and a wall thickness of 0.5\,mm surrounding the fiber. The source position was sampled uniformly in the water volume. The tritium $\beta$ emission energy distribution was taken from Ref. \cite{Mertens2015}. The results are presented in Figure \ref{FIG:2}-\ref{FIG:3b}.  In Figure \ref{FIG:2} the presented distributions were normalized to their own maximum for visualization proposes. It presents the sampled energy distributions for the tritium $\beta$ decay (blue histogram), obtained from the energy distribution given in Ref.\cite{Mertens2015} (red line). The red histogram represents the distribution of the initial energy of the electrons released in the water that are able to reach the fiber and deposit energy in it.  The distribution presents a peak centered at 10 keV, with a broad range from 1\,keV to the maximum energy emission ($Q\,=\,$18\,keV). From the comparison of the distributions it can be observed that the most of low energy electrons, where the tritium $\beta$ emission distribution has it maximum, will not reach the fiber and thus explaining the challenge of tritium in water detection.
 
 Figure \ref{FIG:3b} presents the histogram of the distance between the $\beta$ emission position in water and the fiber surface for the events that deposit energy in the fiber. A narrow peak for small distances is observed. For better visualization the data between 0 and 5\,$\mu$m was re-binned and displayed in the up-left inset of the figure. From the number of entries in the histograms we calculate that the region until 5$\mu$m contains 99.4\% of the total events. It indicates that the majority of electrons that deposit energy in the fibers are generated at a distance shorter than 5\,$\mu$m from the fiber surface. This result allowed us, in order to reduce computing time, to limit the source thickness (water volume around the fiber) to a 5\,$\mu$m water layer. The $\beta$ interaction efficiency ($\epsilon_{\beta}$ - number of events that deposits energy in the fiber normalized to the total number of events generated in the source volume) was calculated presenting a value of 5\%. These results are in agreement with the values presented in Refs. \cite{Rathnakaran2000,Berthold1999} showing that the detector efficiency depends on the sensitive area in contact with water. This value is very low when considering the number of tritium decays for the required activity of 100\,Bq/L in a real-time measurement. The way to increase the detection efficiency in this case is to increase the number and/or the length of the fibers.
 
 \subsection{Light yield and Birks' coefficient}
\label{section:optical}
\begin{figure}
	\centering
		\includegraphics[width=\columnwidth]{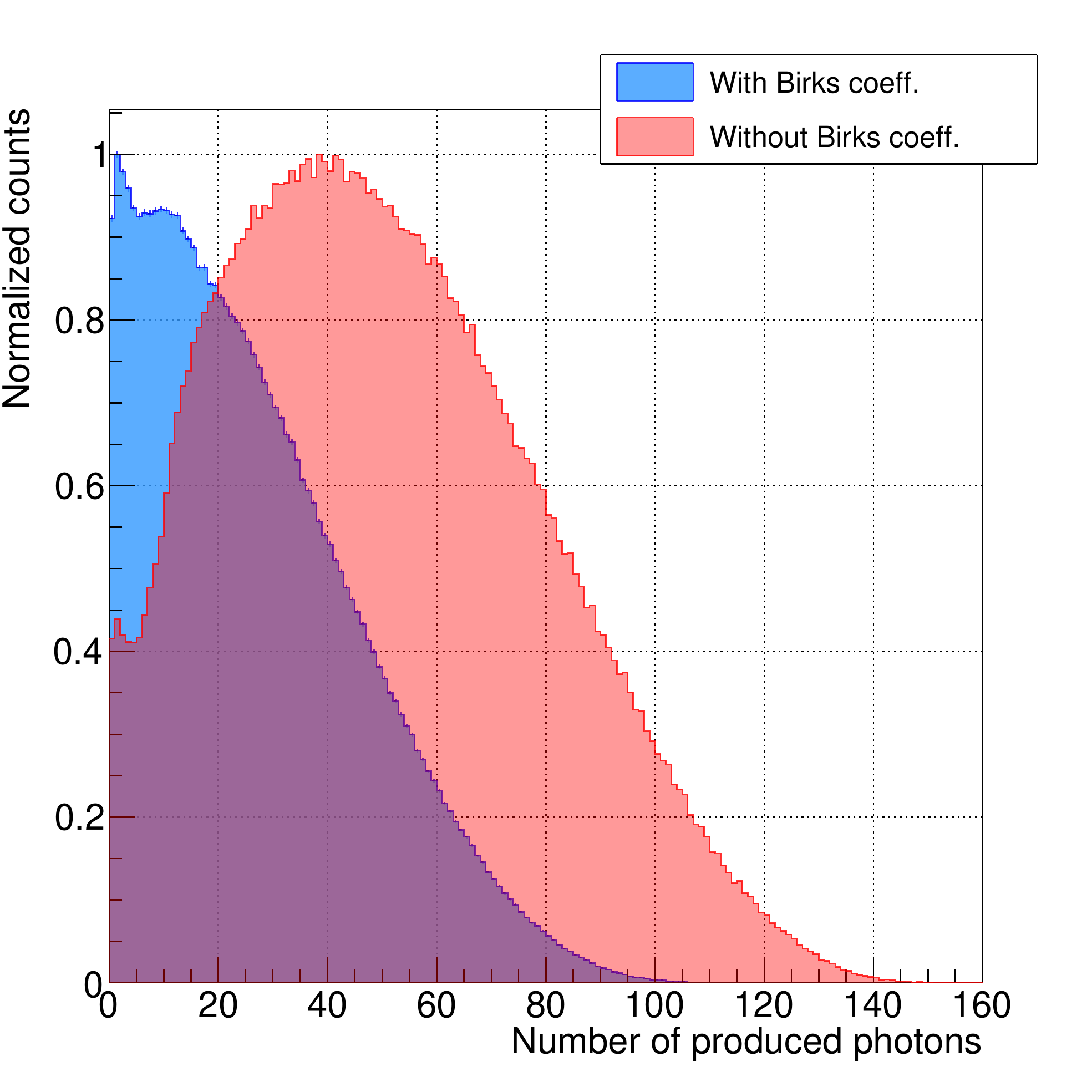}
	\caption{Distribution of scintillation photons produced by the tritium decays in water when interacting with the fiber. Blue: Birks' coefficient considered as 0.126\,mm/MeV. Red:Birks' coefficient considered as 0\,mm/MeV.}
	\label{FIG:4}
\end{figure}

\begin{sloppypar}
In these calculations the scintillation yield of 8000\,photons/MeV stated by the manufacturer for Minimum Ionizing Particles (MIPs) was assumed \cite{SaintGobain}. This value is however one of the main uncertainty in these calculations as the energy region of interest is orders of magnitude lower than that of MIPs. Due to the lack of low energy data for scintillation production, we assumed the number of scintillation photons to be proportional to the deposited energy in the fibers, with the correction for the light quenching, i.e., the Birks' coefficient. 
\end{sloppypar}
\noindent The Birks' coefficient is a correction factor for the scintillators light output per unit of path length ($\frac{dL}{dx}$) \cite{knoll}:
\begin{ceqn}
\begin{align}
  \frac{dL}{dx}=S\frac{\frac{dE}{dx}}{1+k_{B}\frac{dE}{dx}}
\end{align}
\end{ceqn}
\noindent where, $S$ is the scintillation efficiency, $k_{B}$ is Birks' coefficient (material dependent) and $\frac{dE}{dx}$ is the energy loss of the particle per unit of path length. We assumed once again the value for MIPs in polystyrene based scintillators where the  Birks' coefficient is $k_{B}=0.126\,\textrm{mm/MeV}$ \cite{Leverington2011}. This assumption could fail due to a non linear relation between energy deposition and light yield  \cite{Laursonn2016} and to the low energy of tritium electrons compared to MIPs. To investigate the importance of Birks' coefficient, we simulated the light yield taking for $k_B$ either a null value or the MIPs value presenting the results in Figure \ref{FIG:4}. For the case in which Birks' coefficient was considered as zero (red distribution) a peak of 40\,photons produced per $\beta$ particle interacting in the fiber is found in a wide distribution ranging from 1 to 140\,photons. The maximum number of photons value is the expected one by considering the fiber light yield of 8000\,photons/MeV for the maximum $\beta$ emission energy from tritium (18\,keV). When the Birk's coefficient is considered, blue distribution, the light yield reduction is observed: the distribution of the number of photons peaks at $\approx\,$10\,photons and extends up to $\approx\,$110\,photons. 

\section{Full detector simulation}
\begin{figure}
	\centering
		\includegraphics[width=\columnwidth]{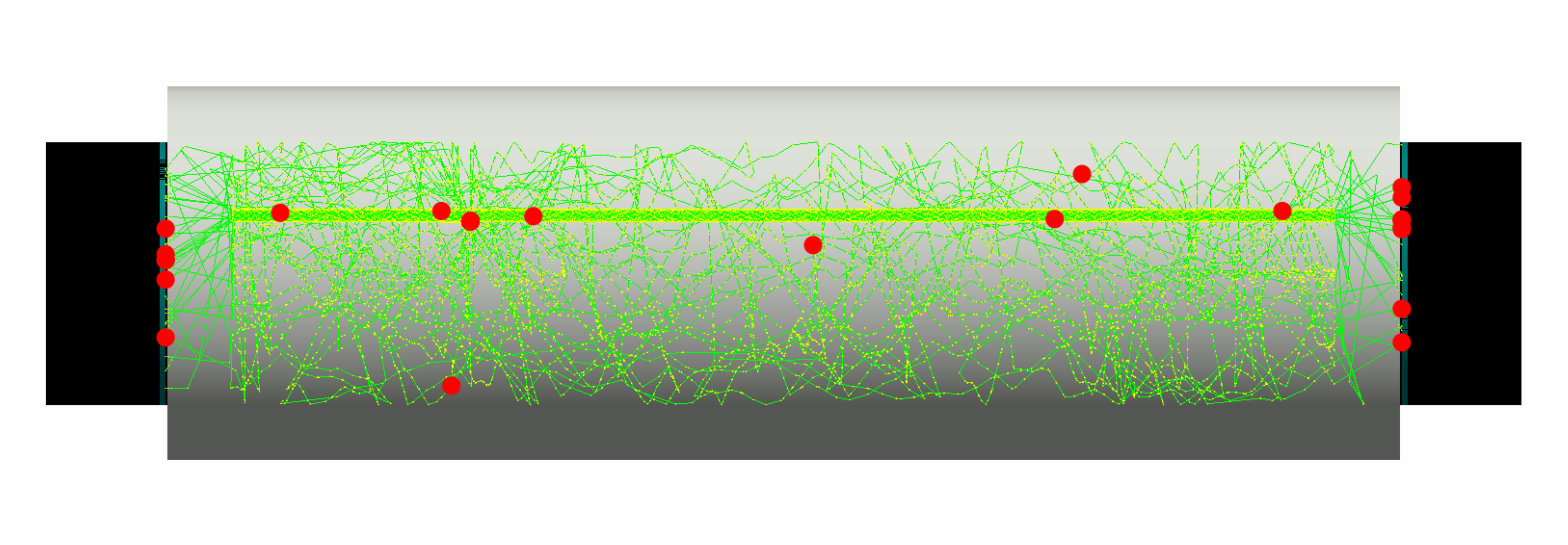}
	\caption{Example of detected events where the red dots are the photons detection positions (in PMTs - black) or optical photon absorption (in the fiber and water regions). Green lines are the optical photons paths}
	\label{FIG:5}
\end{figure}

For the full detector simulation, a 20\,cm and a 1\,m long cells were considered. 340 fibers of 2\,mm diameter surrounded by a 5\,$\mu$m layer of tritiated water were distributed inside the teflon tube of 43\,mm internal diameter fitting the 2" PMTs. The PMTs were optically coupled to the PMMA windows by a 0.5\,mm thick layer of optical silicon grease. An example of an event is shown in Figure \ref{FIG:5} (obtained from GEANT4). In the figure, PMTs are displayed in black at the extremities of the teflon tube (grey). The green lines are the photons paths that, when terminated with a red dot in the PMTs indicate a detected photon. The fiber where the $\beta$ interaction has occurred is clearly distinguished from the other ones by the high density of photons paths. The other photon paths are created by photons not trapped in the fiber critical angle, which travel in water and undergo reflections in the teflon tube internal wall. Red dots along the fiber and in the water represents the photon absorption position by the materials.
 
 \subsection{The fiber length and the modular system}
\begin{figure}
	\centering
		\includegraphics[width=\columnwidth]{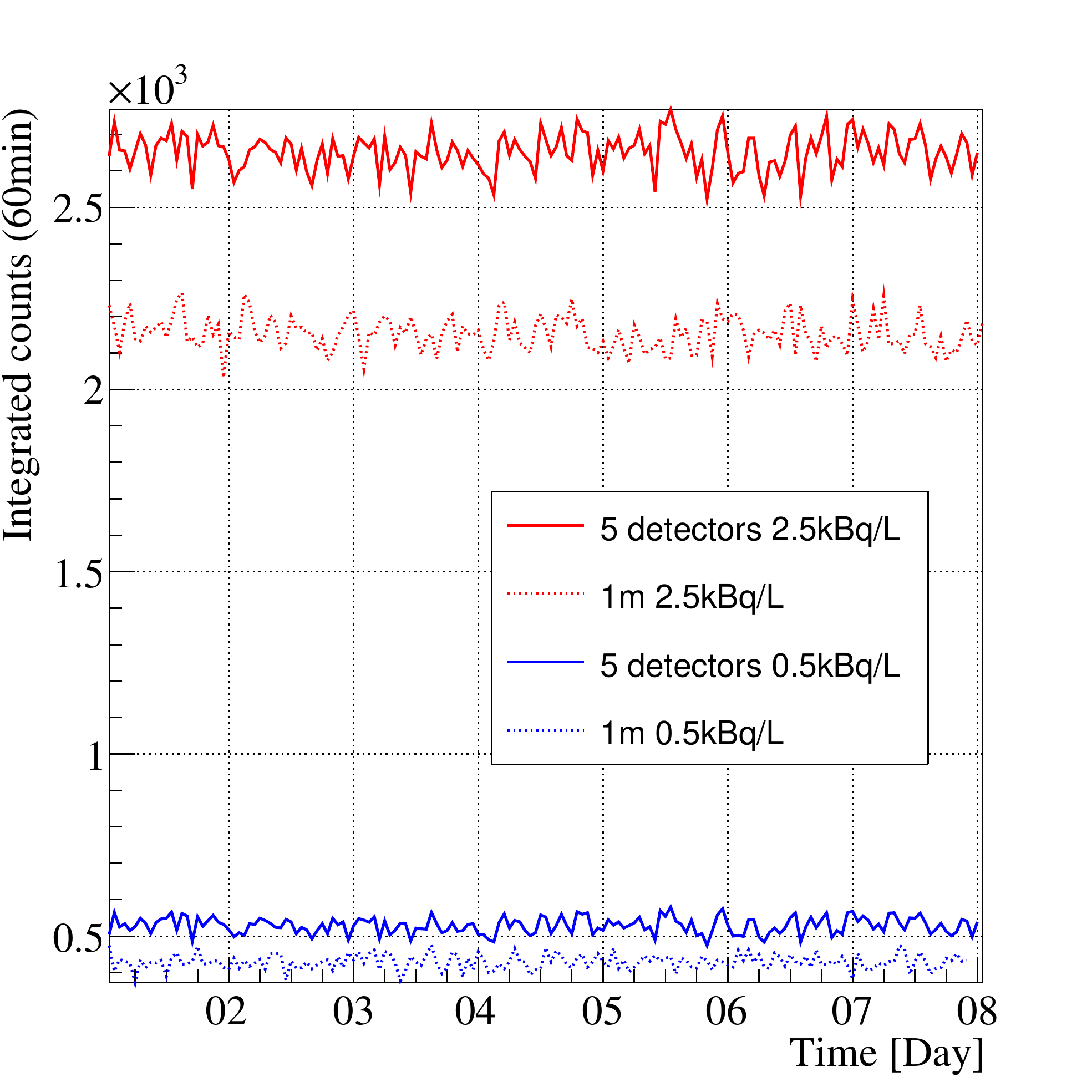}
	\caption{Comparison of the number of coincidences in the detector per hour for a 1\,m long detector and 5 detectors of 20\,cm length.}
	\label{FIG:7}
\end{figure}

 From the results presented in section \ref{source} the $\beta$ interaction efficiency depends on the sensitive surface area in contact with water. Two approaches to increase the surface area are foreseen: Longer fibers (1\,m) or a modular system composed by multiple cells containing short fibers (18\,cm). The advantage of the first approach, 1\,m long fibers is the decrease of the detector price by reducing the number of required PMTs. The second approach, for the same surface area, requires a system of 5 cells and 10 PMTs. On the other side, the advantages of a modular system composed by smaller detectors is the scalability depending on the required sensitivity, a lower photon absorption by the fibers and consequently a higher detection efficiency. Two tritium activities were considered: 0.5\,kBq/L and 2.5\,kBq/L for one week of simulated data taking using a 60\,min counting time. By considering the single-photon detection capability of the PMTs, a detection coincidence is counted when both PMTs have at least one single photon detected. A Birks' coefficient of $k_{B}=0.126\,\textrm{mm/MeV}$ was considered. The results are presented in Figure \ref{FIG:7} where blue and red lines correspond to the simulated activities of 0.5\,kBq/L and 2.5\,kBq/L respectively. Solid lines are the data for the 5 cells (20\,cm long) while dashed lines are the data for the 1\,m long detector. The results indicate a relative increase of the counting rate of about 25\% when short fibers are used, due to the lower photon absorption by the fibers in this case.

\begin{figure}
	\centering
		\includegraphics[width=\columnwidth]{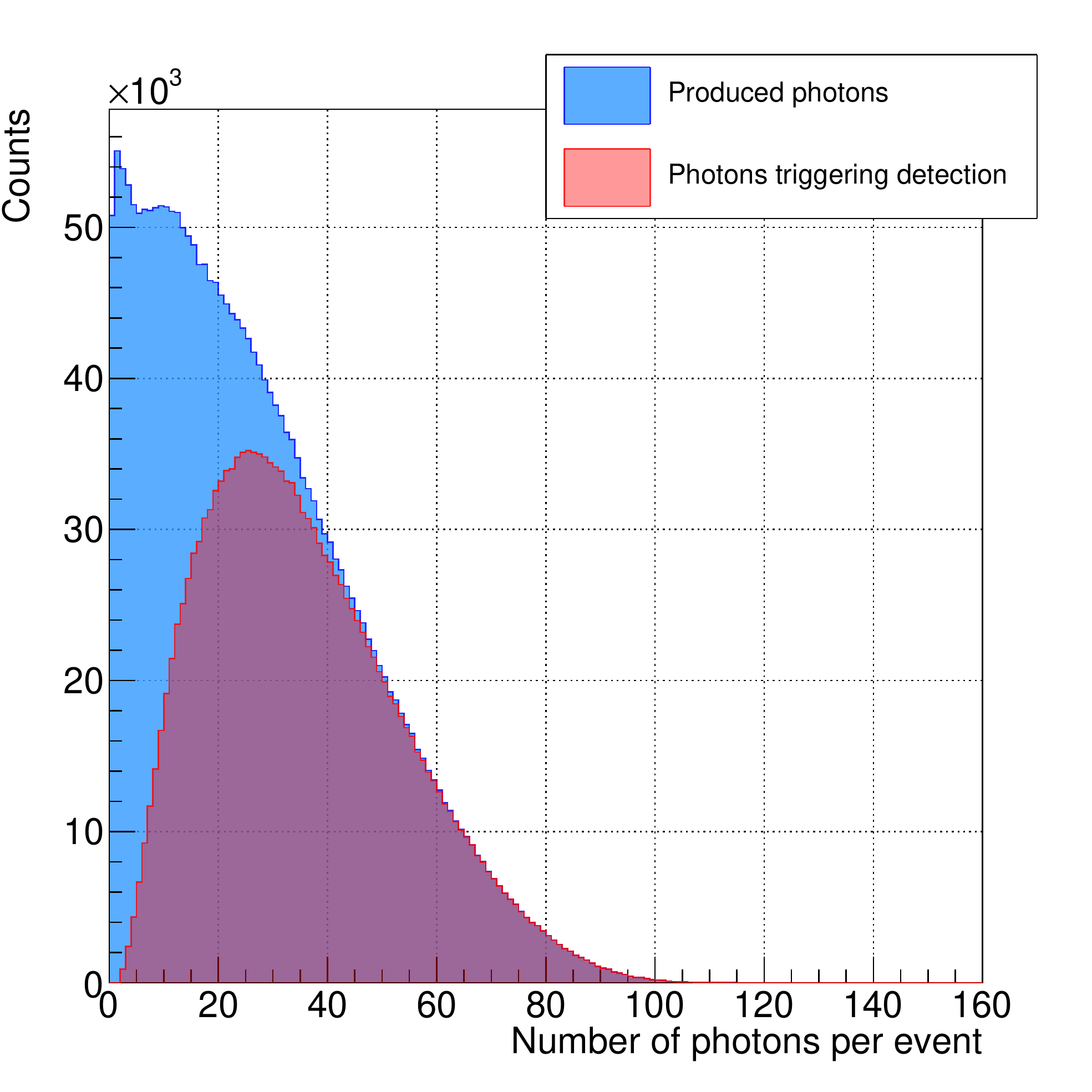}
	\caption{Histograms of the number of produced photons (blue), and the same data when the event is detected by the coincidence both PTMs (red).}
	\label{FIG:6}
\end{figure}

In Figure \ref{FIG:6} the distribution of the number of produced photons per event is shown in blue color, while in red color is the distribution of  the same events when detected in coincidence, obtained in the case of 18\,cm length fibers.
The results reveal that the coincidence detection is peaked for events which produce 25\,photons. The fast decreases on the left side of the peak is  due to the PMTs quantum efficiency and optical losses. On the right side of the peak the distribution matches the blue histogram which indicates that in this region, all the produced events are detected. The coincidence detection efficiency ($\epsilon_{coin}$) is obtained by taking the ratio between the integral of both distributions: $\epsilon_{coin})=$\,67\,\%.
By combining the results presented in section \ref{source}, where a 2\,mm fiber surrounded by a 5\,$\mu$m tritiated water layer presents a $\beta$ interaction efficiency $\epsilon_{\beta}\approx$5\,\% and the $\epsilon_{coin}=$\,67\,\% when Birk's coefficient is considered, a single fiber detection efficiency $\epsilon_{fiber}\approx$\,3.3\,\% is achieved, where $\epsilon_{fiber}=\epsilon_{\beta}\times\epsilon_{coin}$. By considering a bunch of fibers the system detection efficiency will present the same value for a single fiber but, the probability to detect a tritium decay increases proportionally to the number of fibers due to the increase of the water volume sensed.


\subsection{Studies on the integration (counting) time}
\label{section:counting}

\begin{figure}
	\centering
		\includegraphics[width=\columnwidth]{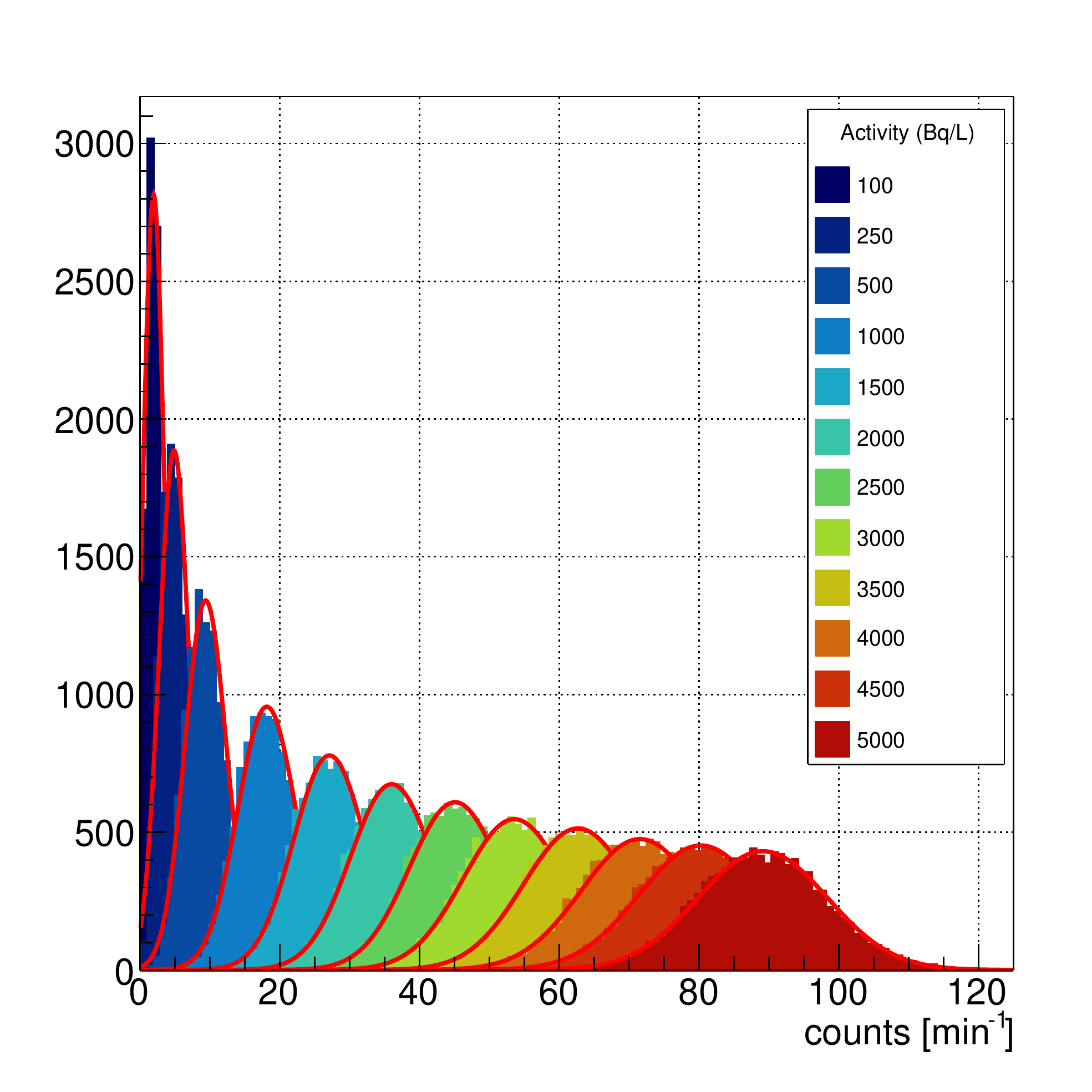}
	\caption{Histograms for the counting rate during a 1\,min integration time for several tritium activities in water. Red lines are Gaussian functions fitted to the different activity data set}
	\label{FIG:8b}
\end{figure}

\begin{figure}
	\centering
		\includegraphics[width=\columnwidth]{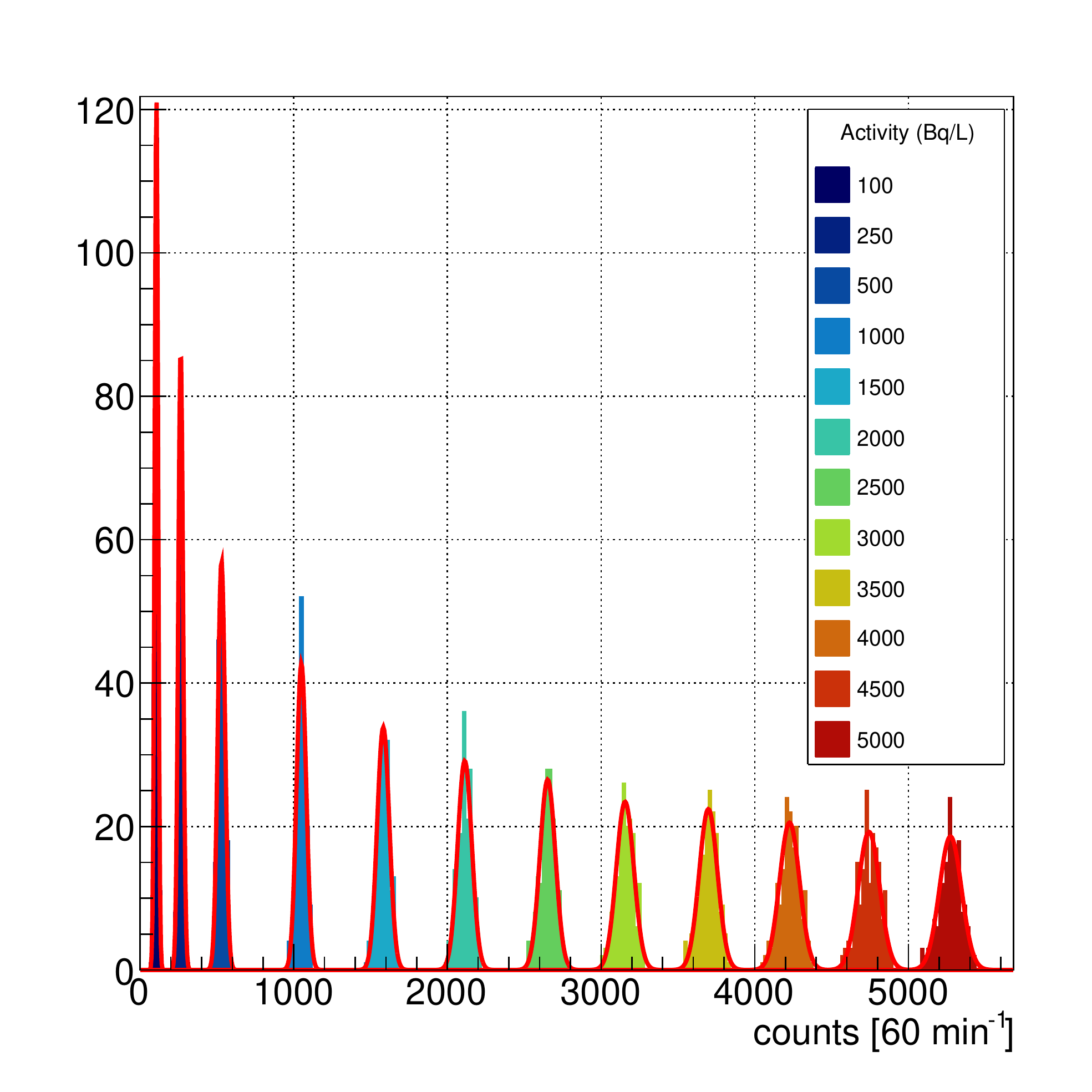}
	\caption{Histograms for the counting rate during a 60\,min integration time for several tritium activities in water. Red lines are Gaussian functions fitted to the different activity data set}
	\label{FIG:9b}
\end{figure}

\begin{figure}
	\centering
		\includegraphics[width=\columnwidth]{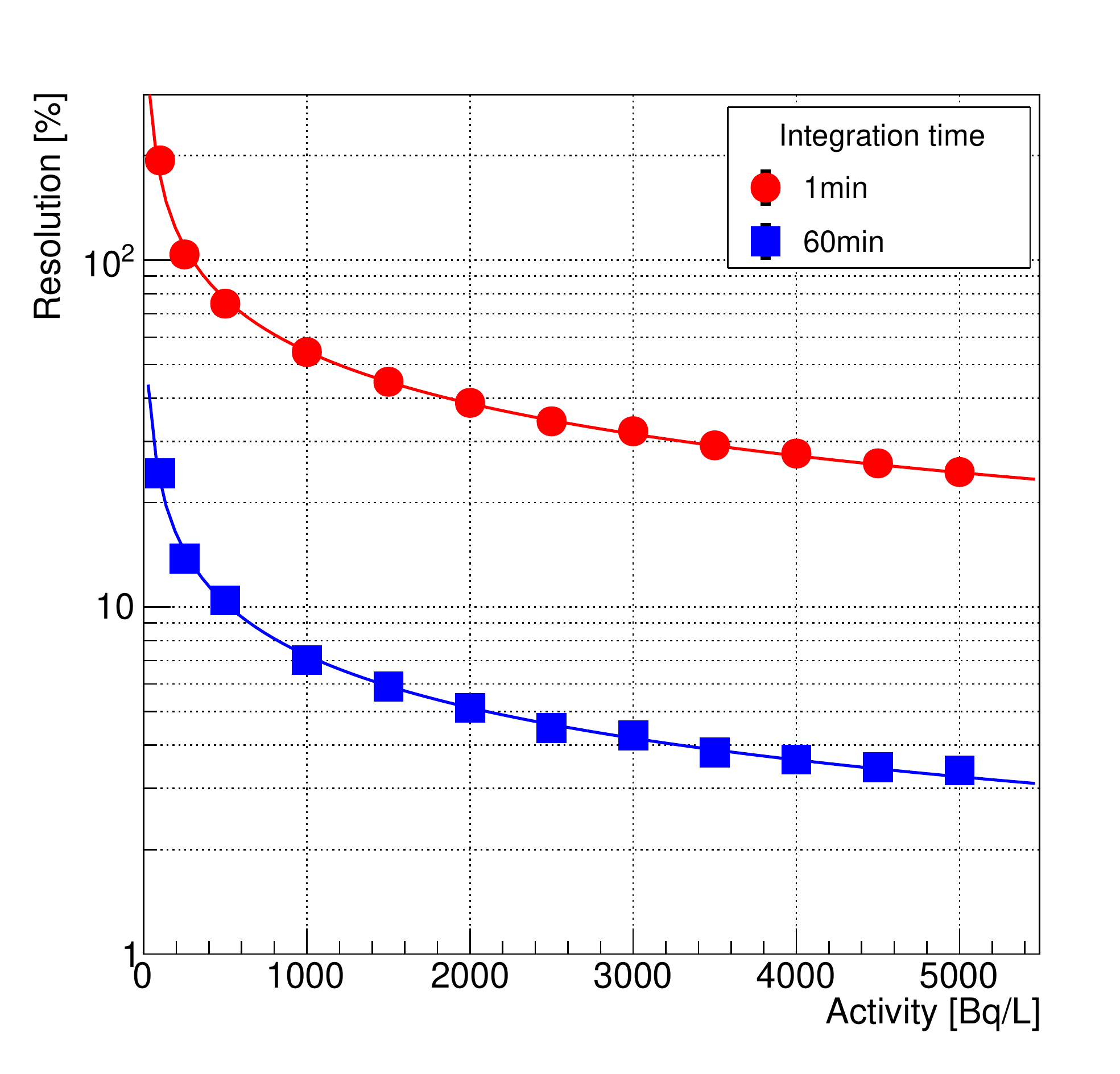}
	\caption{Resolution as a function of the activity for the two considered integration time. The lines correspond to the fitted function $f(x)=C(1/\sqrt{x})$. Error bars are smaller than the marker size}
	\label{FIG:10}
\end{figure}

In order to study the effect of the integration time in the system sensitivity two different analysis were carried out. The data was obtained for the 5 detectors setup and several source activities ranging from 100\,Bq/L to 5\,kBq/L for a week of simulated data taking. In order to get close to "real-time" measurement an integration time of 1\,min was used. The results are presented in Figure \ref{FIG:8b}. Broad distributions were obtained showing the impossibility to distinguish close activity values due to overlap of activities spaced less than $\pm$500\,Bq/L. The broad distributions are due to the large fluctuations from the low $\beta$ interaction efficiency resulting in measurements, i.e, counting with very low statistics and thus suffering from high statistical errors.
In order to decrease the statistical fluctuations the data was analyzed by using 60\,min integration time (what may be denominated as a "quasi-real" time monitoring) being the results presented in Figure \ref{FIG:9b}. It can be observed that the fluctuations are significantly reduced by increasing the integration time. In this case, for the 5 cells setup, all simulated activities can be clearly distinguished even for activities as low as 100 and 250\,Bq/L.
From the data in Figures \ref{FIG:8b} and \ref{FIG:9b} it is possible to extract the resolution (eq. \ref{EQ:2}) of the "measured" activities by fitting a Gaussian function to each data set.
\begin{ceqn}
\begin{align}
Resolution=\frac{FWHM}{centroid}\times100\,(\%)
\label{EQ:2}
\end{align}
\end{ceqn}
\noindent Figure \ref{FIG:10} presents the obtained resolution values for 1 and 60\,min integration time for the simulated tritium activities (dots) while the lines are fitted functions of the expected statistical behavior as function of the measured quantity ($N$):
\begin{ceqn}
\begin{align}
Resolution\propto\frac{1}{\sqrt{N}}
\label{eq:3}
\end{align}
\end{ceqn}
\noindent It can be observed that the resolution for a 60\,min integration is much better when compared to the values obtained for 1\,min integration time. For 60\,min integration time, resolution values ranging from 24.3\% (24\,Bq/L) to 3.4\% (170\,Bq/L) are obtained for the simulated activities of 100\,Bq/L and 5\,kBq/L, respectively.

\subsection{Details on the construction of a prototype}
For the construction of the tritium monitor here described two major concerns must be considered: The radioactive background and the water contaminants. The radioactive background is characterized by the natural and cosmic ray background. For the natural background reduction the use of a lead shield is mandatory (passive shielding) while for the cosmic ray background an anti-coincidence veto (active shielding) must be used. The water contaminants can be distinguished radiative particles and algae/sediments. Radioactive particles can trigger fake event detection while sediments and algae can progressively deposit on the fibers surface disabling the detector sensitivity to low  energy beta particles. To avoid the water contaminants the detector must be operated after a water cleaning system which must filter and de-ionize the water.\\
In parallel to this work 2 prototypes modules based on PMTs and SiPMs, have been produced, tested in laboratory and \emph{in situ} at the water discharge channel from Almaraz power plant dam. In the facility, a lead shield and a cleaning water system which provides water with conductivity < 10\,$\mu$S/cm are installed\cite{Azevedo2018}. The prototypes have been constantly monitored being the detection efficiency limited by the cosmic background. These results will be reported soon in a future publication, including the electronics design, the data acquisition and the slow-control of the detector system.

\section{Conclusions}
In this work we have simulated and studied a modular detection system for real-time monitoring of tritium in water. Due to the low $\beta$ emission energy in the liquid medium a single fiber of 2\,mm diameter can just detect decays occurring at distances shorter than 5\,$\mu$m from its surface with an interaction efficiency of 5\%. From those events interacting with the fiber, 65\% of them will trigger a coincidence between both PMTs resulting in a system detection efficiency around 3.3\%. Due to the low interaction efficiency the sensitivity of the detector for low tritium activities must be increased by increasing the detection area, i.e., the number of fibers. We have shown the advantage on the use of shorter fibers and a modular system which can be scaled according to the desired sensitivity. Uncertainties in the calculations, namely the light yield and Birks' coefficient for low energy $\beta$ particles were addressed.
The results presented in this work show the possibility to build a quasi real-time (60 min integration time) detector for in-water tritium monitoring, with sensitivity to 100\,Bq/L (24\,Bq/L resolution), based on 5 cells, each containing around 340 scintillating fibers of 2\,mm diameter and 18\,cm length.

\section{Acknowledgements}
\begin{sloppypar}
This work was supported by the INTERREG-SUDOE program through the project TRITIUM - SOE1/P4/E0214.
C.D.R. Azevedo was supported by Portuguese national funds (OE), through FCT - Funda\c{c}\~{a}o para a Ci\^{e}ncia e a Tecnologia , I.P., in the scope of the Law 57/2017, of July 19.
Part of this work was also developed within the scope of the project I3N, UIDB/50025/2020 \& UIDP/50025/2020, financed by national funds through the FCT/MEC.
\end{sloppypar}





\end{document}